\documentstyle[preprint,aps]{revtex}

\begin{document}
\draft

\preprint{SLAC-PUB-6335}

\title{
A RELATIVISTIC CONSTITUENT QUARK MODEL
}
\author{Felix Schlumpf}
\address{
Stanford Linear Accelerator Center\\
Stanford University, Stanford, California 94309
}
\date{\today}
\maketitle

\begin{abstract}
We investigate the predictive power of a relativistic quark model
formulated on the light-front. The nucleon electromagnetic form factors,
the semileptonic weak decays of the hyperons and the magnetic moments of
both baryon octet and decuplet are calculated and found to be in excellent
agreement with experiment.
\end{abstract}


\narrowtext

We construct a relativistic constituent quark model consisting of a
radial wave function which is spherically symmetric and invariant under
permutations times a spin-isospin wave function which is uniquely
determined by symmetry requirements [1]. We apply SU(6) symmetry to the rest
frame spinors and boost them to the light-front with a Wigner (Melosh)
rotation. The three-quark wave functions so constructed are eigenfunctions
of mass and spin operators. Eigenfunctions of the four-momentum, which
transform irreducibly under the Poincar\'e group, are obtained from the
mass eigenfunctions using light-front symmetry. The current-density
operator of the constituent quarks is assumed to be that of Dirac point
particles. For the momentum-space wave function a simple function of the
invariant mass $M_0$ is assumed. The invariant mass $M_0$ can be written as
\begin{equation}
M_0^2 = \sum_{i=1}^3 \frac{\vec k_{\perp i}^2 +m_i^2}{x_i} ,
\end{equation}
where we used the longitudinal momentum fractions $x_i=p_i^+/P^+$
($P$ and $p_i$ are the nucleon and quark momenta, respectively, with
$P^+=P_0+P_z$). The internal momentum variables $k_i$ are given by
$k_i=p_i-x_i P$ with $\sum \vec k_{\perp i}=0$ and $\sum x_i=1$. We
choose the following momentum wave functions
\begin{eqnarray}
\phi_H & \sim & \exp (-M_0^2/2\beta^2),
	\label{h} \\
\phi_P & \sim & (1+M_0^2/\beta^2)^{-p}.
	\label{p}
\end{eqnarray}

In Figure~1 the anomalous magnetic moment of the proton $F_2(0)$ is
plotted against the radius $R_1^2=-6F'_1(0)$. Figure~1 shows that the result
for $Q^2=0$ is {\em independent} of the wave function chosen. The only
parameters are:
\begin{itemize}
	\item  The constituent quark  mass $m_i$.

	\item  The scale parameter $\beta$.
\end{itemize}
For small values of $Q^2$, the two wave functions in Eqs.~(\ref{h}) and
(\ref{p}) still give the same results (Fig.~2). Only for very large
momentum transfer $Q^2$ can we see a drastic difference between the
different wave functions as shown in Figure~3.

The parameters of the model are fixed by fitting some of the experimental
values [1]. The nucleon form factors can be calculated for low, medium and
high momentum transfer in excellent agreement with experiment [2]. Figure~3
shows the proton form factor $G_M(Q^2)$ up to more than 30~GeV$^2$.
The broken line gives the form factor calculated with a
conventionally used wave function. That is the reason why it was believed
that the relativistic constituent quark model breaks down at
about 2 GeV$^2$. At
intermediate energies, $G_M$ and $G_E$ for the neutron and $G_M$ for the
proton are in agreement with recent SLAC experiments [3]. A recent pion
bremsstrahlung analysis [4] gives a ratio $\mu(\Delta^{++})/\mu(p)=1.62
\pm 0.18$, much lower than the nonrelativistic quark model value 2, but
in agreement with our value 1.69 [5]. The magnetic moments of the nucleons
and hyperons and the semileptonic decays of the baryon octet are also
described very well with the same parameters [1].

\acknowledgments

It is a pleasure to thank Stan Brodsky for stimulating discussions.
This work was supported in part by the Schweizerischer Nationalfonds and
in part by the Department of Energy, contract DE-AC03-76SF00515.

\begin{figure}
\caption{The anomalous magnetic moment $F_2(0)$ of the proton as a
function of $M_p R_1$: continuous line, pole type wave function; broken
line, gaussian wave function. The experimental value is given by the
cross. Our model is independent of the wave function for $Q^2=0$.}
\end{figure}

\begin{figure}
\caption{The proton form factor $F_{1p}(Q^2)$. The line code is the same
as in the previous figure.}
\end{figure}

\begin{figure}
\caption{The proton form factor $G_M(Q^2)$: continuous line is the present
analysis; broken line gives form factor calculated with a conventionally
used wave function. The relativistic constituent quark model does not
break down at 2 GeV$^2$, it is even valid up to more than 30 GeV$^2$.}
\end{figure}

\begin{figure}
\caption{The axial vector form factor $g_1(K^2=-Q^2)$ for the
neutron-proton weak decay.}
\end{figure}


\begin{thebibliography}{10}

\bibitem{1}  F. Schlumpf, Phys. Rev. D  {\bf 47}, 4114 (1993).

\bibitem{2} F. Schlumpf, SLAC-PUB-5968 (1992) to be published in J. Phys. G;
SLAC-PUB-6050 (1993) to be published in Mod. Phys. Lett. A.

\bibitem{3}  P. E. Bosted {\it et al.}, Phys. Rev. Lett. {\bf 68}, 3841 (1992);
A. Lung {\it et al.}, Phys. Rev. Lett. {\bf 70}, 718 (1993).

\bibitem{4}  A.~Bosshard {\it et~al.}, Phys. Rev. D {\bf 44}, 1962 (1991).

\bibitem{5} F. Schlumpf, SLAC-PUB-6218 (1993) to be published  in Phys. Rev. D.

\end{thebibliography}
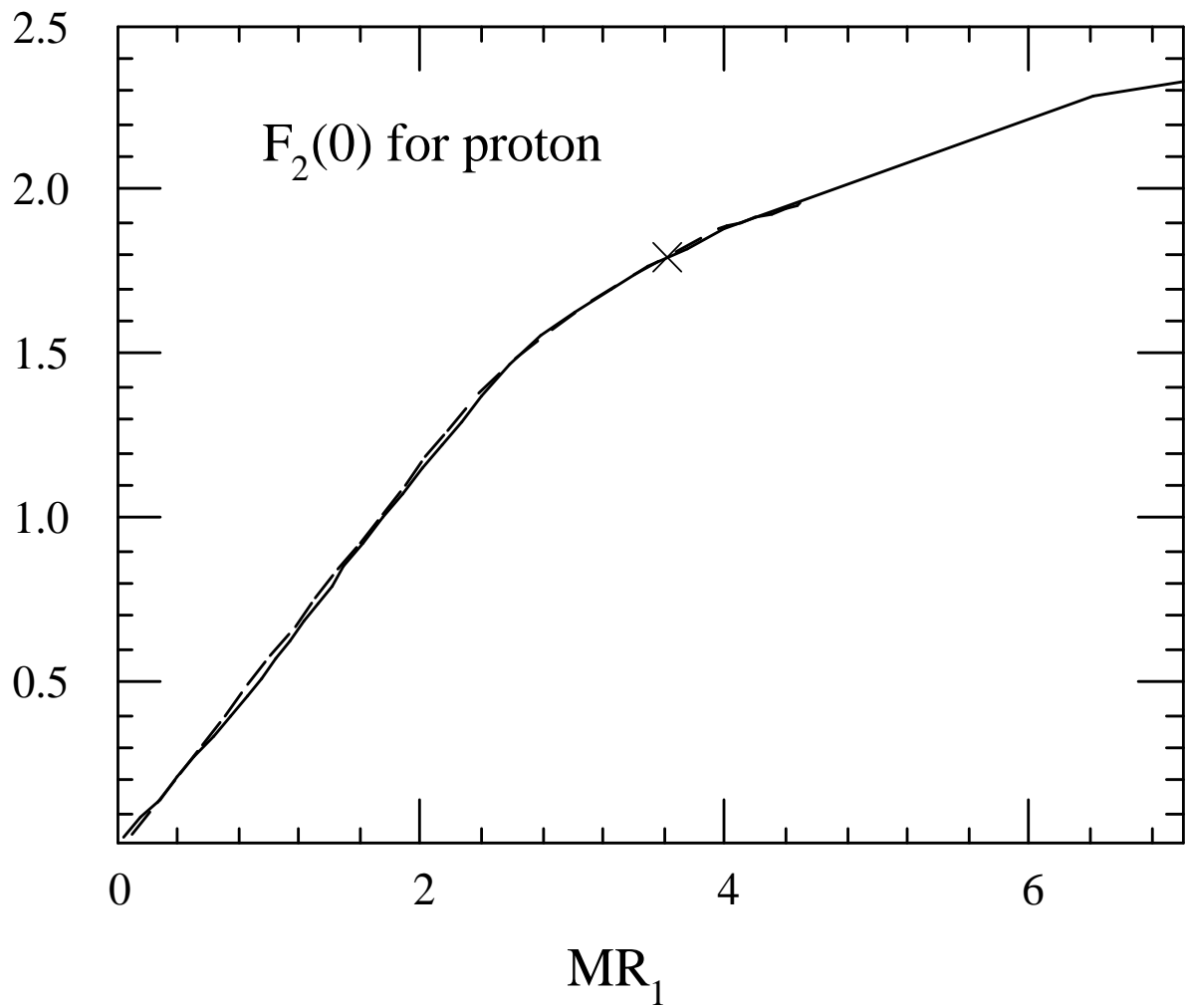
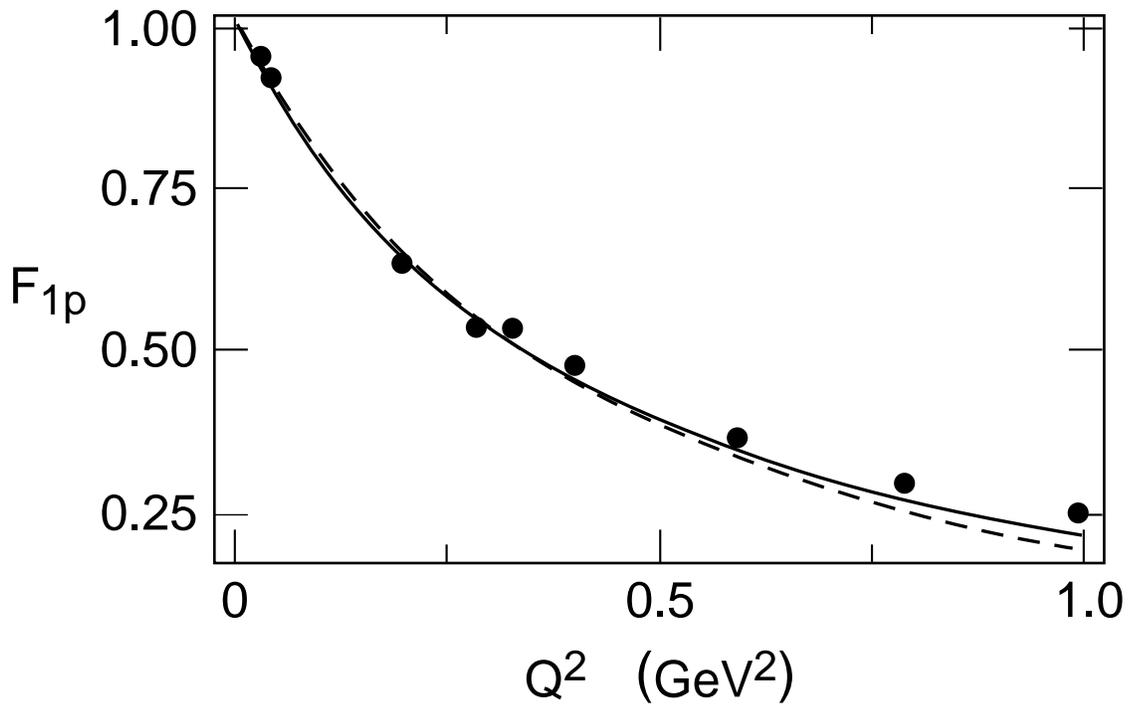
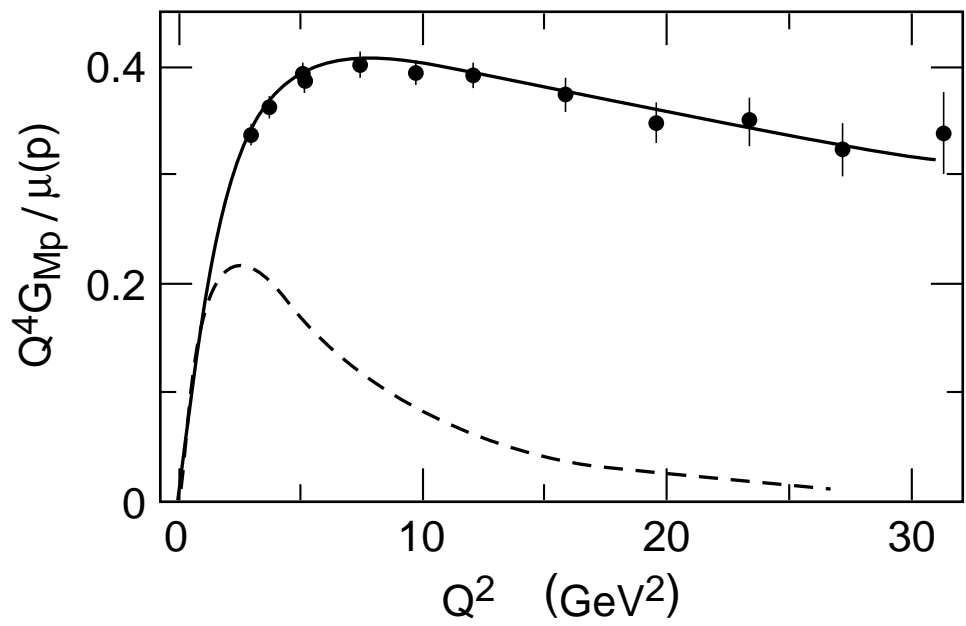
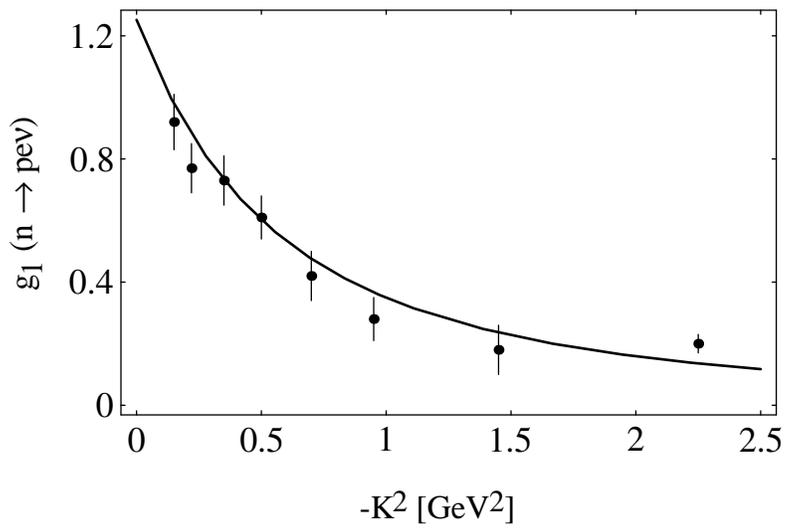
\end{document}